\documentclass[conference]{IEEEtran}

\usepackage{amsmath, amssymb, amsthm, mathtools}
\usepackage{relsize, paralist, hyperref, xcolor, balance, setspace}
\usepackage[T1]{fontenc}

\newtheorem{theorem}{Theorem}
\newtheorem{proposition}[theorem]{Proposition}
\newtheorem{definition}[theorem]{Definition}

\newcommand{ \C }{ \bs{C} }
\newcommand{ \myF }{ \mathbb{F} }

\newcommand{ \myG }{ \mathcal G }
\newcommand{ \myK }{ \mathcal K }
\newcommand{ \myP }{ \mathcal P }

\newcommand{ \myU }{ \mathcal U }
\newcommand{ \myX }{ \mathcal X }
\newcommand{ \myY }{ \mathcal Y }
\newcommand{ \Z }{ \mathbb{Z} }
\newcommand{ \N }{ \mathbb{N} }
\newcommand{ \rank }{ \operatorname{rank} }
\newcommand{ \myarrow }{ \stackrel{\sml{\myK}}{\rightsquigarrow} }
\newcommand{ \sml }[1]{ \mathsmaller{#1} }
\newcommand{ \bs }[1]{ \boldsymbol{#1} }

\begin{document}

\title{\huge Optimal Error-Detecting Codes for General Asymmetric Channels via Sperner Theory}

\author{%
\IEEEauthorblockN{Mladen~Kova\v{c}evi\'c and Dejan~Vukobratovi\'{c}}
\IEEEauthorblockA{Faculty of Technical Sciences, University of Novi Sad, Serbia\\
                  Emails: kmladen@uns.ac.rs, dejanv@uns.ac.rs}
}

\maketitle

\begin{abstract}
Several communication models that are of relevance in practice are
asymmetric in the way they act on the transmitted ``objects''.
Examples include channels in which the amplitudes of the transmitted
pulses can only be decreased, channels in which the symbols can only
be deleted, channels in which non-zero symbols can only be shifted to
the right (e.g., timing channels), subspace channels in which the
dimension of the transmitted vector space can only be reduced, unordered
storage channels in which the cardinality of the stored (multi)set can
only be reduced, etc.
We introduce a formal definition of an asymmetric channel as a channel
whose action induces a partial order on the set of all possible inputs,
and show that this definition captures all the above examples.
Such a general approach allows one to treat all these different models
in a unified way, and to obtain a characterization of optimal
error-detecting codes for many interesting asymmetric channels by using
Sperner theory.
\end{abstract}%


\section{Introduction}
\label{sec:intro}

Several important channel models possess an intrinsic asymmetry in the
way they act on the transmitted ``objects''.
A classical example is the binary $ \mathsf{Z} $-channel in which the
transmitted $ 1 $'s may be received as $ 0 $'s, but not vice versa.
In this article we formalize the notion of an asymmetric channel by using
order theory, and illustrate that the given definition captures this and
many more examples.
Our main goals are the following:
\begin{inparaenum}
\item[1)]
to introduce a framework that enables one to treat many different kinds
of asymmetric channels in a unified way, and
\item[2)]
to demonstrate its usefulness and meaningfulness through examples.
In particular, the usefulness of the framework is illustrated by describing
\emph{optimal} error-detecting codes for a broad class of asymmetric channels
(for all channel parameters), a result that follows from Kleitman's theorem
on posets satisfying the so-called LYM inequality.
\end{inparaenum}

\subsection{Communication channels}
\label{sec:channels}

\begin{definition}
\label{def:channel}
Let $ \myX, \myY $ be nonempty sets.
A communication channel on $ (\myX, \myY) $ is a subset
$ \myK \subseteq \myX \times \myY $ satisfying\linebreak
$ \forall x \in \myX \; \exists y \in \myY  \; (x,y) \in \myK $ and
$ \forall y \in \myY \; \exists x \in \myX  \; (x,y) \in \myK $.
We also use the notation $ {x \myarrow y} $, or simply $ x \rightsquigarrow y $
when there is no risk of confusion, for $ (x,y) \in \myK $.

For a given channel $ \myK \subseteq \myX \times \myY $, we define its
dual channel as $ \myK^\textnormal{d} = \{ (y, x) : (x, y) \in \myK \} $.
\end{definition}

Note that we describe communication channels purely in combinatorial terms,
as \emph{relations} in Cartesian products $ \myX \times \myY $.\linebreak
Here $ \myX $ is thought of as the set of all possible inputs, and $ \myY $
as the set of all possible outputs of the channel.
The \pagebreak expression  $ x \rightsquigarrow y $ means that the input $ x $ can
produce the output $ y $ with positive probability.
We do not assign particular values of probabilities to each pair
$ (x,y) \in \myK $ as they are irrelevant for the problems that we
intend to discuss.

\subsection{Partially ordered sets}
\label{sec:posets}

In what follows, we shall use several notions from order theory, so we
recall the basics here \cite{engel, stanley}.

A partially ordered set (or poset) is a set $ \myU $ together with a
relation $ \preceq $ satisfying, for all $ x, y, z \in \myU $:
\begin{inparaenum}
\item[1)]
reflexivity:
$ x \preceq x $,
\item[2)]
asymmetry (or antisymmetry):
if $ x \preceq y $ and $ y \preceq x $, then $ x = y $,
\item[3)]
transitivity:
if $ x \preceq y $ and $ y \preceq z $, then $ x \preceq z $.
\end{inparaenum}
Two elements $ x, y \in \myU $ are said to be comparable if either
$ x \preceq y $ or $ y \preceq x $.
They are said to be incomparable otherwise.
A chain in a poset $ (\myU, \preceq) $ is a subset of
$ \myU $ in which any two elements are comparable.
An antichain is a subset of $ \myU $ any two distinct elements of which
are incomparable.

A function $ \rho: \myU \to \mathbb{N} $ is called a rank function if
$ \rho(y) = \rho(x) + 1 $ whenever $ y $ covers $ x $, meaning that
$ x \preceq y $ and there is no $ y' \in \myU $ such that $ x \preceq y' \preceq y $.
A poset with a rank function is called graded.
In a graded poset with rank function $ \rho $ we denote
$ \myU_{[\underline{\ell}, \overline{\ell}]} =
  \{ x \in \myU : \underline{\ell} \leqslant \rho(x) \leqslant \overline{\ell} \} $,
and we also write $ \myU_\ell = \myU_{[\ell,\ell]} $ (here the rank function
$ \rho $ is omitted from the notation as it is usually understood from the
context).
Hence, $ \myU = \bigcup_\ell \myU_\ell $.
A graded poset is said to have Sperner property if $ \myU_\ell $ is an antichain
of maximum cardinality in $ (\myU, \preceq) $, for some $ \ell $.
A poset is called rank-unimodal if the sequence $ |\myU_\ell| $ is unimodal
(i.e., an increasing function of $ \ell $ when $ \ell \leqslant \ell' $,
and decreasing when $ \ell \geqslant \ell' $, for some $ \ell' $).

We say that a graded poset $ (\myU, \preceq) $ possesses the LYM
(Lubell--Yamamoto--Meshalkin) property \cite{kleitman} if there exists a
nonempty list of maximal chains such that, for any $ \ell $, each of the
elements of rank $ \ell $ appear in the same number of chains.
In other words, if there are $ L $ chains in the list, then each element
of rank $ \ell $ appears in $ L/|\myU_\ell| $ of the chains.
We shall call a poset \emph{normal} if it satisfies the LYM property,
see \cite[Sec.~4.5 and Thm 4.5.1]{engel}.
A simple sufficient condition for a poset to be normal is that it be
regular \cite[Cor.~4.5.2]{engel}, i.e., that both the number of elements that
cover $ x $ and the number of elements that are covered by $ x $ depend only
on the rank of $ x $.

In Section \ref{sec:examples} we shall see that many standard examples of
posets, including the Boolean lattice, the subspace lattice, the Young's
lattice, chain products, etc., arise naturally in the analysis of
communications channels.
\pagebreak

\section{General asymmetric channels and\\error-detecting codes}
\label{sec:asymmetric}

In this section we give a formal definition of asymmetric channels and
the corresponding codes which unifies and generalizes many scenarios
analyzed in the literature.
We assume hereafter that the sets of all possible channel inputs and all
possible channels outputs are equal, $ \myX = \myY $.

For a very broad class of communication channels, the relation
$ \rightsquigarrow $ is reflexive, i.e., $ x \rightsquigarrow x $ (any
channel input can be received unimpaired, in case there is no noise), and
transitive, i.e., if $ x \rightsquigarrow y $ and $ y \rightsquigarrow z $,
then $ x \rightsquigarrow z $ (if there is a noise pattern that transforms
$ x $ into $ y $, and a noise pattern that transforms $ y $ into $ z $,
then there is a noise pattern  -- a combination of the two -- that transforms
$ x $ into $ z $).
Given such a channel, we say that it is \emph{asymmetric} if the relation
$ \rightsquigarrow $ is asymmetric, i.e., if $ x \rightsquigarrow y $,
$ x \neq y $, implies that $ y \not\rightsquigarrow x $.
In other words, we call a channel asymmetric if the channel action induces
a partial order on the space of all inputs $ \myX $.

\begin{definition}
\label{def:asymmetric}
  A communication channel $ \myK \subseteq \myX^2 $ is said to be asymmetric
if $ (\myX, \stackrel{\sml{\myK}}{\rightsquigarrow}) $ is a partially ordered
set.
We say that such a channel is * if the poset
$ (\myX, \stackrel{\sml{\myK}}{\rightsquigarrow}) $ is *, where * stands for
an arbitrary property a poset may have (e.g., graded, Sperner, normal, etc.).
\end{definition}

Many asymmetric channels that arise in practice, including all the examples
mentioned in this paper, are graded as there are natural rank functions that
may be assigned to them.
For a graded channel $ \myK $, we denote by
$ \myK_{[\underline{\ell}, \overline{\ell}]} =
  \myK \cap \big( \myX_{[\underline{\ell}, \overline{\ell}]} \big)^{\!2} $
its natural restriction to inputs of rank $ \underline{\ell}, \ldots, \overline{\ell} $.

\begin{definition}
\label{def:edc}
We say that $ \bs{C} \subseteq \myX $ is a code detecting up to $ t $ errors
in a graded asymmetric channel $ \myK \subseteq \myX^2 $ if, for all $ x, y \in \C $,
\begin{align}
\label{eq:detectgen}
  x \myarrow y  \; \land \;  x \neq y  \quad \Rightarrow \quad  | \rank(x) - \rank(y) | > t .
\end{align}
We say that $ \bs{C} \subseteq \myX $ detects \emph{all} error patterns in
an asymmetric channel $ \myK \subseteq \myX^2 $ if, for all $ x, y \in \C $,
\begin{align}
\label{eq:detectgen2}
  x \myarrow y  \quad \Rightarrow \quad  x = y .
\end{align}
\end{definition}

For graded channels, the condition \eqref{eq:detectgen2} is satisfied
if and only if the condition \eqref{eq:detectgen} holds for any $ t $.

In words, $ \bs{C} $ detects all error patterns in a given asymmetric channel
if no element of $ \C $ can produce another element of $ \C $ at the channel
output.
If this is the case, the receiver will easily recognize whenever the transmission
is erroneous because the received object is not going to be a valid codeword
which could have been transmitted.
Yet another way of saying that $ \C $ detects all error patterns is the following.

\begin{proposition}
\label{thm:edc}
  $ \C \subseteq \myX $ detects all error patterns in an asymmetric channel
$ \myK \subseteq \myX^2 $ if and only if $ \C $ is an antichain in the
corresponding poset $ (\myX, \stackrel{\sml{\myK}}{\rightsquigarrow}) $.
\end{proposition}

A simple example of an antichain, and hence a code detecting all error
patterns in a graded asymmetric channel, is the level set $ \myX_\ell $,
for an arbitrary $ \ell $.
\pagebreak
\begin{definition}
\label{def:optimal}
  We say that $ \C \subseteq \myX $ is an optimal code detecting up to
$ t $ errors (resp. all error patterns) in a graded asymmetric channel
$ \myK \subseteq \myX^2 $ if there is no code of cardinality larger than
$ |\C| $ that satisfies \eqref{eq:detectgen} (resp. \eqref{eq:detectgen2}).
\end{definition}

Hence, an optimal code detecting all error patterns in an asymmetric channel
$ \myK \subseteq \myX^2 $ is an antichain of maximum cardinality in the poset
$ (\myX, \stackrel{\sml{\myK}}{\rightsquigarrow}) $.
Channels in which the code $ \myX_\ell $ is optimal, for some $ \ell $, are
called Sperner channels.
All channels treated in this paper are Sperner.

An example of an error-detecting code, of which the code $ \myX_\ell $ is
a special case (obtained for $ t \to \infty $), is given in the following
proposition.

\begin{proposition}
\label{thm:tedc}
  Let $ \myK \subseteq \myX^2 $ be a graded asymmetric channel, and
$ (\ell_n)_n $ a sequence of integers satisfying $ \ell_n - \ell_{n-1} > t $,
$ \forall n $.
The code $ \C = \bigcup_{n} \myX_{\ell_n} $ detects up to $ t $ errors in $ \myK $.
\end{proposition}

If the channel is normal, an optimal code detecting up to $ t $ errors is
of the form given in Proposition \ref{thm:tedc}.
We state this fact for channels which are additionally rank-unimodal, as
this is the case that is most common.

\begin{theorem}
\label{thm:optimal}
  Let $ \myK \subseteq \myX^2 $ be a normal rank-unimodal asymmetric channel.
The maximum cardinality of a code detecting up to $ t $ errors in
$ \myK_{[\underline{\ell}, \overline{\ell}]} $ is given by
\begin{equation}
\label{eq:maxsumgen}
  \max_{m}  \sum^{\overline{\ell}}_{\substack{ \ell=\underline{\ell} \\ \ell \, \equiv \, m \; (\operatorname{mod}\, t+1) } }  |\myX_\ell| .
\end{equation}
\end{theorem}
\begin{IEEEproof}
This is essentially a restatement of the result of Kleitman~\cite{kleitman}
(see also \cite[Cor.~4.5.4]{engel}) which states
that, in a finite normal poset $ ( \myU, \preceq ) $, the largest cardinality
of a family $ \C \subseteq \myU $ having the property that, for all distinct
$ x, y \in \C $, $ x \preceq y $ implies that $ \rank(y) - \rank(x) > t $, is
$ \max_F \sum_{x \in F} |\myU_{\rank(x)}| $.
The maximum here is taken over all chains $ F = \{x_1, x_2, \ldots, x_c\} $
satisfying $ x_1 \preceq x_2 \preceq \cdots \preceq x_c $ and
$ \rank(x_{i+1}) - \rank(x_i) > t $ for $ i = 1, 2, \ldots, c-1 $, and all
$ c = 1, 2, \ldots $.
If the poset $ ( \myU, \preceq ) $ is in addition rank-unimodal, then it is
easy to see that the maximum is attained for a chain $ F $ satisfying
$ \rank(x_{i+1}) - \rank(x_i) = t + 1 $ for $ i = 1, 2, \ldots, c-1 $, and
that the maximum cardinality of a family $ \C $ having the stated property
can therefore be written in the simpler form
\begin{equation}
\label{eq:maxsumgen2}
  \max_{m}  \sum_{\ell \, \equiv \, m  \; (\operatorname{mod}\, t+1)}  |\myU_\ell| .
\end{equation}
Finally, \eqref{eq:maxsumgen} follows by recalling that the restriction
$ ( \myU_{[\underline{\ell}, \overline{\ell}]}, \preceq ) $ of a normal
poset $ ( \myU, \preceq ) $ is normal \cite[Prop. 4.5.3]{engel}.
\end{IEEEproof}
\vspace{2mm}

We note that an optimal value of $ m $ in \eqref{eq:maxsumgen} can be
determined explicitly in many concrete examples (see Section~\ref{sec:examples}).

We conclude this section with the following claim which enables one to
directly apply the results pertaining to a given asymmetric channel to
its dual.

\begin{proposition}
\label{thm:dual}
  A channel $ \myK \subseteq \myX^2 $ is asymmetric if and only if its dual
$ \myK^\textnormal{d} $ is asymmetric.
A code $ \bs{C} \subseteq \myX $ detects up to $ t $ errors in $ \myK $ if
and only if it detects up to $ t $ errors in $ \myK^\textnormal{d} $.
\end{proposition}

\section{Examples}
\label{sec:examples}

In this section we list several examples of communication channels that
have been analyzed in the literature in different contexts and that are
asymmetric in the sense of Definition \ref{def:asymmetric}.
For each of them, a characterization of optimal error-detecting codes is
given based on Theorem \ref{thm:optimal}.

\subsection{Codes in power sets}
\label{sec:subset}

Consider a communication channel with $ \myX = \myY = 2^{\{1,\ldots,n\}} $
and with $ A \rightsquigarrow B $ if and only if $ B \subseteq A $, where
$ A, B \subseteq \{1, \ldots, n\} $.
Codes defined in the power set $ 2^{\{1,\ldots,n\}} $ were proposed in
\cite{gadouleau+goupil2, kovacevic+vukobratovic_clet} for error control
in networks that randomly reorder the transmitted packets (where the set
$ \{1,\ldots,n\} $ is identified with the set of all possible packets), and
are also of interest in scenarios where data is written in an unordered way,
such as DNA-based data storage systems \cite{lenz}.
Our additional assumption here is that the received set is always a subset
of the transmitted set, i.e., the noise is represented by ``set reductions''.
These kinds of errors may be thought of as consequences of packet losses/deletions.
Namely, if $ t $ packets from the transmitted set $ A $ are lost in the
channel, then the received set $ B $ will be a subset of $ A $ of cardinality
$ |A| - t $.
We are interested in codes that are able to detect up to $ t $ packet
deletions, i.e., codes having the property that if $ B \subsetneq A $,
$ |A| - |B| \leqslant t $, then $ A $ and $ B $ cannot both be codewords.

It is easy to see that the above channel is asymmetric in the sense of
Definition \ref{def:asymmetric}; the ``asymmetry'' in this model is
reflected in the fact that the cardinality of the transmitted set can
only be reduced.
The poset $ ( \myX, \rightsquigarrow ) $ is the so-called Boolean lattice
\cite[Ex.~1.3.1]{engel}.
The rank function associated with it is the set cardinality:
$ \rank(A) = |A| $, for any $ A \subseteq \{1, \ldots, n\} $.
This poset is rank-unimodal, with $ |\myX_\ell| = \binom{n}{\ell} $, and
normal \cite[Ex.~4.6.1]{engel}.
By applying Theorem~\ref{thm:optimal} we then obtain the maximum cardinality
of a code $ \C \subseteq 2^{\{1,\ldots,n\}} $ detecting up to $ t $ deletions.
Furthermore, an optimal value of $ m $ in \eqref{eq:maxsumgen} can be
found explicitly in this case.
This claim was first stated by Katona~\cite{katona} in a different terminology.

\begin{theorem}
\label{thm:subset}
  The maximum cardinality of a code $ \C \subseteq 2^{\{1,\ldots,n\}} $
detecting up to $ t $ deletions is
\begin{equation}
\label{eq:maxsumsets}
  \sum^n_{\substack{ \ell=0 \\ \ell \, \equiv \, \lfloor \frac{n}{2} \rfloor  \; (\operatorname{mod}\, t+1) } }
	  \binom{n}{\ell}
\end{equation}
\end{theorem}

Setting $ t \to \infty $ (in fact, $ t > \lceil n/2 \rceil $ is sufficient),
we conclude that the maximum cardinality of a code detecting any number
of deletions is $ \binom{n}{\lfloor n/2 \rfloor} = \binom{n}{\lceil n/2 \rceil} $.
This is a restatement of the well-known Sperner's theorem \cite{sperner},
\cite[Thm 1.1.1]{engel}.

For the above channel, its dual (see Definition~\ref{def:channel}) is the
channel with $ \myX = 2^{\{1, \ldots, n\}} $ in which $ A \rightsquigarrow B $
if and only if $ B \supseteq A $.
This kind of noise, ``set augmentation'', may be thought of as a consequence
of packet insertions.
Proposition~\ref{thm:dual} implies that the expression in \eqref{eq:maxsumsets}
is also the maximum cardinality of a code $ \C \subseteq \myX $ detecting up
to $ t $ insertions.

\subsection{Codes in the space of multisets}
\label{sec:multiset}

A natural generalization of the model from the previous subsection,
also motivated by unordered storage or random permutation channels,
is obtained by allowing repetitions of symbols, i.e., by allowing
the codewords to be \emph{multisets} over a given alphabet \cite{multiset}. 

A multiset $ A $ over $ \{1, \ldots, n\} $ can be uniquely described by
its multiplicity vector $ \mu_A = (\mu_A(1), \ldots, \mu_A(n)) \in \N^n $,
where $ \N = \{0, 1, \ldots\} $.
Here $ \mu_A(i) $ is the number of occurrences of the symbol $ i \in \{1, \ldots, n\} $
in $ A $.
We again consider the deletion channel in which $ A \rightsquigarrow B $
if and only if $ B \subseteq A $ or, equivalently, if $ \mu_B \leqslant \mu_A $
(coordinate wise).

If we agree to use the multiplicity vector representation of multisets, we may
take $ \myX = \myY = \N^n $.
The channel just described is asymmetric in the
sense of Definition~\ref{def:asymmetric}.
The rank function associated with the poset $ {(\myX, \rightsquigarrow)} $ is
the multiset cardinality: $ \rank(A) = \sum_{i=1}^n \mu_A(i) $.
We have $ |\myX_\ell| = \binom{\ell + n - 1}{n - 1} $.

The following claim is a multiset analog of Theorem~\ref{thm:subset}.

\begin{theorem}
\label{thm:multiset}
  The maximum cardinality of a code $ \C \subseteq \myX_{[\underline{\ell}, \overline{\ell}]} $,
$ \myX = \N^n $, detecting up to $ t $ deletions is
\begin{align}
\label{eq:Mcodesize}
  \sum^{\lfloor \frac{\overline{\ell} - \underline{\ell}}{t+1} \rfloor}_{i=0}
	 \binom{\overline{\ell} - i (t+1) + n - 1}{n - 1} .
\end{align}
\end{theorem}
\begin{IEEEproof}
The poset $ (\myX, \rightsquigarrow) $ is normal as it is a product of
chains \cite[Ex.~4.6.1]{engel}.
We can therefore apply Theorem~\ref{thm:optimal}.\linebreak
Furthermore, since $ |\myX_\ell| = \binom{\ell + n - 1}{n - 1} $ is a
monotonically increasing function of $ \ell $, the optimal choice of $ m $
in \eqref{eq:maxsumgen} is $ \overline{\ell} $, which implies \eqref{eq:Mcodesize}.
\end{IEEEproof}
\vspace{2mm}

The dual channel is the channel in which $ A \rightsquigarrow B $ if and
only if $ B \supseteq A $, i.e., $ \mu_B \geqslant \mu_A $.
These kinds of errors -- multiset augmentations -- may be caused by
insertions or duplications.

\subsection{Codes for the binary $ \mathsf{Z} $-channel and its generalizations}
\label{sec:Z}

Another interpretation of Katona's theorem \cite{katona} in the coding-theoretic
context, easily deduced by identifying subsets of $ \{1, \ldots, n\} $
with sequences in $ \{0, 1\}^n $, is the following: the expression in
\eqref{eq:maxsumsets} is the maximum size of a binary code of length
$ n $ detecting up to $ t $ asymmetric errors, i.e., errors of the form
$ 1 \to 0 $ \cite{borden}.
By using Kleitman's result \cite{kleitman}, Borden~\cite{borden} also
generalized this statement and described optimal codes over arbitrary
alphabets detecting $ t $ asymmetric errors.
(Error control problems in these kinds of channels have been studied
quite extensively; see, e.g., \cite{blaum, bose+rao}.)

To describe the channel in more precise terms, we take
$ \myX = \myY = \{0, 1, \ldots, a-1\}^n $ and we let
$ (x_1, \ldots, x_n) \rightsquigarrow (y_1, \ldots, y_n) $ if and only
if $ y_i \leqslant x_i $ for all $ i = 1, \ldots, n $.
This channel is asymmetric and the poset $ (\myX, \rightsquigarrow) $
is normal \cite[Ex.~4.6.1]{engel}.
The appropriate rank function here is the Manhattan weight:
$ \rank(x_1, \ldots, x_n) = \sum_{i=1}^n x_i $.
In the binary case ($ {a = 2} $), this channel is called the $ \mathsf{Z} $-channel
and the Manhattan weight coincides with the Hamming weight.

Let $ c(N, M, \ell) $ denote the number of \emph{compositions} of the
number $ \ell $ with $ M $ non-negative parts, each part being $ \leqslant\! N $
\cite[Sec.~4.2]{andrews}.
In other words, $ c(N, M, \ell) $ is the number of vectors from
$ \{0, 1, \ldots, N-1\}^M $ having Manhattan weight $ \ell $.
Restricted integer compositions are well-studied objects;
for an explicit expression for $ c(N, M, \ell) $, see \cite[p.~307]{stanley}.

\begin{theorem}[Borden \cite{borden}]
\label{thm:Z}
  The maximum cardinality of a code $ \C \subseteq \{0, 1, \ldots, a-1\}^n $
detecting up to $ t $ asymmetric errors is
\begin{align}
\label{eq:Zcode}
  \sum^{n(a-1)}_{\substack{ \ell=0 \\ \ell \, \equiv \, \lfloor \frac{n(a-1)}{2} \rfloor \; (\operatorname{mod}\, t+1) }}
	 c(a-1, n, \ell) .
\end{align}
\end{theorem}

The channel dual to the one described above is the channel in which
$ (x_1, \ldots, x_n) \rightsquigarrow (y_1, \ldots, y_n) $ if and only
if $ y_i \geqslant x_i $ for all $ i = 1, \ldots, n $.

\subsection{Subspace codes}
\label{sec:subspace}

Let $ \myF_q $ denote the field of $ q $ elements, where $ q $ is a prime
power, and $ \myF_q^n $ an $ n $-dimensional vector space over $ \myF_q $.
Denote by $ \myP_q(n) $ the set of all subspaces of $ \myF_q^n $
(also known as the projective space), and by $ \myG_q(n , \ell) $
the set of all subspaces of dimension $ \ell $ (also known as the Grassmannian).
The cardinality of $ \myG_q(n , \ell) $ is expressed through
the $ q $-binomial (or Gaussian) coefficients \cite[Ch.~24]{vanlint+wilson}:
\begin{align}
\label{eq:gcoeff}
  \left| \myG_q(n , \ell) \right|
   = \binom{n}{\ell}_{\! q}
	 = \prod_{i=0}^{\ell-1} \frac{ q^{n-i} - 1 }{ q^{\ell-i} - 1 } .
\end{align}
The following well-known properties of $ \binom{n}{\ell}_{\! q} $ will
be useful:
\begin{inparaenum}
\item[1)]
symmetry: $ \binom{n}{\ell}_{\! q} = \binom{n}{n-\ell}_{\! q} $, and
\item[2)]
unimodality: $ \binom{n}{\ell}_{\! q} $ is increasing in $ \ell $ for
$ \ell \leqslant \frac{n}{2} $, and decreasing for $ \ell \geqslant \frac{n}{2} $.
\end{inparaenum}
We use the convention that $ \binom{n}{\ell}_{\! q} = 0 $ when
$ \ell < 0 $ or $ \ell > n $.

Codes in $ \myP_q(n) $ were proposed in \cite{koetter+kschischang} for error
control in networks employing random linear network coding \cite{ho}, in which
case $ \myF_q^n $ corresponds to the set of all length-$ n $ packets (over a
$ q $-ary alphabet) that can be exchanged over the network links.
We consider a channel model in which the only impairments are ``dimension
reductions'', meaning that, for any given transmitted vector space
$ U \subseteq \myF_q^n $, the possible channel outputs are subspaces of $ U $.
These kinds of errors can be caused by packet losses, unfortunate choices
of the coefficients in the performed linear combinations in the network
(resulting in linearly dependent packets at the receiving side), etc.

In the notation introduced earlier, we set $ \myX = \myY = \myP_q(n) $ and
define the channel by: $ U \rightsquigarrow V $ if and only if $ V $ is a
subspace of $ U $.
This channel is asymmetric.
The poset $ (\myX, \rightsquigarrow) $ is the so-called linear lattice
(or the subspace lattice) \cite[Ex.~1.3.9]{engel}.
The rank function associated with it is the dimension of a vector space:
$ \rank(U) = \dim U $, for $ U \in \myP_q(n) $.
We have $ |\myX_\ell| = | \myG_q(n , \ell) | = \binom{n}{\ell}_{\! q} $.

The following statement may be seen as the $ q $-analog \cite[Ch.~24]{vanlint+wilson}
of Katona's theorem \cite{katona}, or of Theorem \ref{thm:subset}.

\begin{theorem}
\label{thm:subspace}
  The maximum cardinality of a code $ \C \subseteq \myP_q(n) $
detecting dimension reductions of up to $ t $ is
\begin{align}
\label{eq:codesize}
  \sum^n_{\substack{ \ell=0 \\ \ell \, \equiv \, \lfloor \frac{n}{2} \rfloor  \; (\operatorname{mod}\, t+1) } }  \binom{n}{\ell}_{\! q} .
\end{align}
\end{theorem}
\begin{IEEEproof}
The poset $ (\myP_q(n), \subseteq) $ is rank-unimodal and normal
\cite[Ex.~4.5.1]{engel} and hence, by Theorem \ref{thm:optimal},
the maximum cardinality of a code detecting dimension reductions
of up to $ t $ can be expressed in the form
\begin{subequations}
\begin{align}
\label{eq:maxsum}
  &\max_{m}  \sum^n_{\substack{ \ell=0 \\ \ell \, \equiv \, m  \; (\operatorname{mod}\, t+1) } }
	  \binom{n}{\ell}_{\! q}  \\
\label{eq:maxsumr}
  &\;\;\;\;= \max_{r \in \{0, 1, \ldots, t\}} \; \sum_{j \in \Z}  \binom{n}{\lfloor \frac{n}{2} \rfloor + r + j(t+1)}_{\! q} .
\end{align}
\end{subequations}
(Expression \eqref{eq:maxsum} was also given in \cite[Thm~7]{ahlswede+aydinian}.)
We need to show that $ m = \lfloor n / 2 \rfloor $ is a maximizer
in \eqref{eq:maxsum} or, equivalently, that $ r = 0 $ is a maximizer in
\eqref{eq:maxsumr}.
Let us assume for simplicity that $ n $ is even; the proof for odd $ n $ is similar.
What we need to prove is that the following expression is non-negative,
for any $ r \in \{1, \ldots, t\} $,
\begin{subequations}
\label{eq:j}
\begin{align}
\nonumber
  &\sum_{j \in \Z}  \binom{n}{\frac{n}{2} + j(t+1)}_{\! q} -
	 \sum_{j \in \Z}  \binom{n}{\frac{n}{2} + r + j(t+1)}_{\! q}  \\
\label{eq:jpos}
	&\;\;= \sum_{j > 0}  \binom{n}{\frac{n}{2} + j(t+1)}_{\! q} - \binom{n}{\frac{n}{2} + r + j(t+1)}_{\! q} +  \\
\label{eq:jzero}
  &\;\;\phantom{=}\ \binom{n}{\frac{n}{2}}_{\! q} - \binom{n}{\frac{n}{2} + r}_{\! q} - \binom{n}{\frac{n}{2} + r - (t+1)}_{\! q} +  \\
\label{eq:jneg}
	&\;\;\phantom{=}\ \sum_{j < 0}  \binom{n}{\frac{n}{2} + j(t+1)}_{\! q} - \binom{n}{\frac{n}{2} + r + (j-1)(t+1)}_{\! q} .
\end{align}
\end{subequations}
Indeed, since the $ q $-binomial coefficients are unimodal and maximized
at $ \ell = n / 2 $, each of the summands in the sums \eqref{eq:jpos} and
\eqref{eq:jneg} is non-negative, and the expression in \eqref{eq:jzero} is
also non-negative because%
\begin{subequations}
\begin{align}
\nonumber
  &\binom{n}{\frac{n}{2}}_{\! q} - \binom{n}{\frac{n}{2} + r}_{\! q} - \binom{n}{\frac{n}{2} + r - (t+1)}_{\! q}  \quad  \\
\label{eq:a1}
	&\;\;\;\;\geqslant  \binom{n}{\frac{n}{2}}_{\! q} - \binom{n}{\frac{n}{2} + 1}_{\! q} - \binom{n}{\frac{n}{2} -1}_{\! q}  \\
\label{eq:a2}
	&\;\;\;\;= \binom{n}{\frac{n}{2}}_{\! q} - 2 \binom{n}{\frac{n}{2} - 1}_{\! q}  \\
\label{eq:a3}
	&\;\;\;\;= \binom{n}{\frac{n}{2}}_{\! q}  \left( 1 - 2 \frac{q^{\frac{n}{2} + 1} - 1}{q^{\frac{n}{2} + 2} - 1} \right)  \\
\label{eq:a4}
	&\;\;\;\;> \binom{n}{\frac{n}{2}}_{\! q}  \left( 1 - 2 \frac{1}{q} \right)  \\
\label{eq:a5}
	&\;\;\;\;\geqslant 0 ,
\end{align}
\end{subequations}
where \eqref{eq:a1} and \eqref{eq:a2} follow from unimodality and symmetry
of $ \binom{n}{\ell}_{\! q} $, \eqref{eq:a3} is obtained by substituting the
definition of $ \binom{n}{\ell}_{\! q} $, \eqref{eq:a4} follows from the fact
that $ \frac{\alpha-1}{\beta-1} < \frac{\alpha}{\beta} $ when $ 1 < \alpha < \beta $,
and \eqref{eq:a5} is due to $ q \geqslant 2 $.
\end{IEEEproof}
\vspace{2mm}

As a special case when $ t \to \infty $ (in fact, $ t > \lceil n/2 \rceil $
is sufficient), we conclude that the maximum cardinality of a code detecting
arbitrary dimension reductions is $ \binom{n}{\lfloor n/2 \rfloor}_{\! q} $.
In other words, $ \myG_q(n, \lfloor n/2 \rfloor) $ is an antichain of maximum
cardinality in the poset $ (\myP_q(n), \subseteq) $ (see Prop.~\ref{thm:edc}).
This is the well-known $ q $-analog of Sperner's theorem \cite[Thm 24.1]{vanlint+wilson}.

The dual channel in this example is the channel in which
$ U \rightsquigarrow V $ if and only if $ U $ is a subspace of $ V $.

\subsection{Codes for deletion and insertion channels}
\label{sec:deletions}

Consider the channel with
$ \myX = \myY = \{0, 1, \ldots, a-1\}^* = \bigcup_{n=0}^\infty \{0, 1, \ldots, a-1\}^n $
in which $ x \rightsquigarrow y $ if and only if $ y $ is a subsequence
of $ x $.
This is the so-called deletion channel in which the output sequence is
produced by deleting some of the symbols of the input sequence.
The channel is asymmetric in the sense of Definition~\ref{def:asymmetric}.
The rank function associated with the poset $ (\myX, \rightsquigarrow) $ is
the sequence length: for any $ x = x_1 \cdots x_\ell $, where $ x_i \in \{0, 1, \ldots, a-1\} $,
$ \rank(x) = \ell $.
We have $ |\myX_\ell| = a^\ell $.

Given that $ \myX $ is infinite, we shall formulate the following
statement for the restriction $ \myX_{[\underline{\ell}, \overline{\ell}]} $,
i.e., under the assumption that only sequences of lengths
$ \underline{\ell}, \ldots, \overline{\ell} $ are allowed as inputs.
This is a reasonable assumption from the practical viewpoint.

\begin{theorem}
The maximum cardinality of a code
$ \C \subseteq \bigcup_{\ell=\underline{\ell}}^{\overline{\ell}} \{0, 1, \ldots, a-1\}^\ell $
detecting up to $ t $ deletions is
\begin{align}
  \sum_{j=0}^{\lfloor \frac{\overline{\ell} - \underline{\ell}}{t+1} \rfloor}
	  a^{\overline{\ell} - j (t+1)} .
\end{align}
\end{theorem}
\begin{IEEEproof}
The poset $ (\myX_{[0, \overline{\ell}]}, \rightsquigarrow) $ is normal.
To see this, note that the list of $ a^{\overline{\ell}} $ maximal chains
of the form $ \epsilon \rightsquigarrow x_1 \rightsquigarrow x_1 x_2
\rightsquigarrow \cdots \rightsquigarrow x_1 x_2 \ldots x_{\overline{\ell}} $,
where $ \epsilon $ is the empty sequence and $ x_i \in \{0, 1, \ldots, a-1\} $,
satisfies the condition that each element of $ \myX_{[0, {\overline{\ell}}]} $
of rank $ \ell $ appears in the same number of chains, namely $ a^{\overline{\ell} - \ell} $
(see Section~\ref{sec:posets}).
The claim now follows by invoking Theorem \ref{thm:optimal} and by using
the fact that $ |\myX_\ell| = a^\ell $ is a monotonically increasing function
of $ \ell $, implying that the optimal choice for $ m $ in \eqref{eq:maxsumgen}
is $ \overline{\ell} $.
\end{IEEEproof}
\vspace{2mm}

The dual channel in this example is the insertion channel in which
$ x \rightsquigarrow y $ if and only if $ x $ is a subsequence of $ y $.

\subsection{Codes for bit-shift and timing channels}
\label{sec:shift}

Let $ \myX = \myY = \{0, 1\}^n $, and let us describe binary sequences by
specifying the positions of $ 1 $'s in them.
More precisely, we identify $ x \in \{0,1\}^n $ with
the integer sequence $ \lambda_x = (\lambda_x(1), \ldots, \lambda_x(w)) $,
where $ \lambda_x(i) $ is the position of the $ i $'th $ 1 $ in $ x $,
and $ w $ is the Hamming weight of $ x $.
This sequence satisfies
$ 1 \leqslant \lambda_x(1) < \lambda_x(2) < \cdots < \lambda_x(w) \leqslant n $.
For example, for $ x = 1 1 0 0 1 0 1 $, $ \lambda_x = (1, 2, 5, 7) $.
In fact, it will be more convenient to use a slightly different description
of a sequence $ x $, namely $ \tilde{\lambda}_x = \lambda_x - (1, 2, \ldots, w) $,
for which it holds that $ 0 \leqslant \tilde{\lambda}_x(1) \leqslant
\tilde{\lambda}_x(2) \leqslant \cdots \leqslant \tilde{\lambda}_x(w) \leqslant n - w $.

Consider a communication model in which each of the $ 1 $'s in the input
sequence may be shifted to the right \cite{shamai+zehavi, kovacevic}.
Such models are also useful for describing timing channels wherein
$ 1 $'s indicate the time slots in which packets have been sent and
shifts of these $ 1 $'s are consequences of packet delays; see for
example \cite{kovacevic+popovski}.
Thus $ x \rightsquigarrow y $ if and only if $ x $ and $ y $ have the
same Hamming weight and $ \lambda_x \leqslant \lambda_y $ (coordinate wise).
Since a necessary condition for $ x \rightsquigarrow y $ is that $ x $ and
$ y $ have the same Hamming weight, we may consider the sets of inputs
$ \{0, 1\}^n_w \equiv \{ x \in \{0, 1\}^n : \sum_{i=1}^n x_i = w\} $
separately, for each $ w = 0, \ldots, n $ (here $ x = x_1 \cdots x_n $).

The above channel is asymmetric.
The poset $ (\{0, 1\}^n_w, \rightsquigarrow) $ is denoted $ L(n-w, w) $
in \cite[Ex.~1.3.13]{engel}.
The rank function on this poset is defined by:
$ \rank(x) = \sum_{i=1}^{w} \tilde{\lambda}_x(i) $, where $ w $ is the
Hamming weight of $ x $.

Let $ p(N, M, \ell) $ denote the number of \emph{partitions} of the number
$ \ell $ into at most $ M $ positive parts, each part being $ \leqslant\! N $
\cite[Sec.~3.2]{andrews}.
These too are very well-studied objects.
An interesting connection between them and the Gaussian coefficients which
we encountered in Section~\ref{sec:subspace} is the following
\cite[Sec.~3.2]{andrews}, \cite[Thm 24.2]{vanlint+wilson}:
\begin{align}
\label{eq:partitions}
  \sum_{\ell=0}^{MN}  p(N, M, \ell) q^\ell = \binom{N+M}{M}_{\! q} .
\end{align}

\begin{theorem}
The maximum cardinality of a code $ \C \subseteq \{0, 1\}^n_w $ detecting
up to $ t $ right-shifts is lower-bounded by
\begin{align}
\label{eq:shift}
  \sum^{w (n-w)}_{\substack{ \ell=0 \\ \ell \, \equiv \, \lfloor \frac{w(n-w)}{2} \rfloor  \; (\operatorname{mod}\, t+1) } }  p(n-w, w, \ell) .
\end{align}
The maximum cardinality of a code $ \C \subseteq \{0, 1\}^n_w $ detecting
all patterns of right-shifts is $ p(n-w, w, \lfloor \frac{w(n-w)}{2} \rfloor) $.
\end{theorem}
\begin{IEEEproof}
The number of elements in $ \{0, 1\}^n_w $ of rank $ \ell $ is $ p(n-w, w, \ell) $.
These numbers are symmetric, $ p(n-w, w, \ell) = p(n-w, w, w(n-w) - \ell) $,
and unimodal, and hence maximized when $ \ell = \lfloor \frac{w(n-w)}{2} \rfloor $
\cite[Thm 3.10]{andrews}.
Furthermore, it follows from \cite[Thm 6.2.10 and Cor.~6.2.1]{engel} that the
poset $ (\{0, 1\}^n_w, \rightsquigarrow) $ is Sperner.
This implies the second statement.
The first statement follows from Proposition \ref{thm:tedc}.
\end{IEEEproof}
\vspace{2mm}

We believe the lower bound in \eqref{eq:shift} is actually the optimal value,
i.e., the maximum cardinality of a code detecting $ t $ right-shifts, but at
present we do not have a proof of this fact.

The dual channel in this example is the channel in which non-zero
symbols may be shifted only to the left.

\section{Conclusion}

As we have seen, order theory is a powerful tool for analyzing asymmetric
channel models, particularly the error detection problem for which an optimal
solution may be obtained in many cases of interest.
Developing the introduced framework further and exploring other applications
and channel models that fit into it is a topic of ongoing investigation.

Note that we have not discussed here error-\emph{correcting} codes in the
posets we encountered.
This is also left for future work (see \cite{firer} for a related study).

\vspace{3mm}
\emph{Acknowledgment:}
This work was supported by European Union's Horizon 2020 research and
innovation programme (Grant Agreement no.\ 856967), and by the
Secretariat for Higher Education and Scientific Research of the Autonomous
Province of Vojvodina (project no.\ 142-451-2686/2021).

\balance

\end{document}